\documentstyle{cupconf}

\ifoldfss
\else
  \ifnfssone
    \newmathalphabet{\mathit}
      \addtoversion{normal}{\mathit}{cmr}{m}{it}
      \addtoversion{bold}{\mathit}{cmr}{bx}{it}
    \newmathalphabet{\mathcal}
      \addtoversion{normal}{\mathcal}{cmsy}{m}{n}
    \else
    \ifnfsstwo
    \fi
  \fi
\fi

\def\hexnumber#1{\ifcase#1 0\or1\or2\or3\or4\or5\or6\or7\or8\or9\or
 A\or B\or C\or D\or E\or F\fi }

\makeatletter
\ifx\CUP@mtlplain@loaded\undefined
\else
\fi
\makeatother
 \makeatletter
 \ifx\CUP@mtlplain@loaded\undefined
   \font\tenbmi=cmmib10 at 10pt
   \font\sevenbmi=cmmib10 at 7pt
   \font\fivebmi=cmmib10 at 5pt

   \newfam\bmifam
   \textfont\bmifam=\tenbmi
   \scriptfont\bmifam=\sevenbmi
   \scriptscriptfont\bmifam=\fivebmi
   
 \fi
 \makeatother
\ifnfsstwo

\fi
\ifnfssone

\fi
\ifoldfss

\fi

\mathchardef\varLambda="0103

\makeatletter
\ifx\CUP@mtlplain@loaded\undefined
\else
\fi
\makeatother
\makeatletter
\ifx\CUP@mtlplain@loaded\undefined
  \font\tenbms=cmbsy10
  \font\sevenbms=cmbsy10 at 7pt
  \font\fivebms=cmbsy10 at 5pt
  \newfam\bmsfam
  \textfont\bmsfam=\tenbms
  \scriptfont\bmsfam=\sevenbms
  \scriptscriptfont\bmsfam=\fivebms

  \edef\bsy@{\hexnumber\bmsfam}
  \mathchardef\bnabla="0\bsy@72
\fi
\makeatother
\title[High redshift ULIRGs]{Deep sub-mm surveys: 
High redshift ULIRGs and the formation of the metal-rich spheroids
}

\author[S. J. Lilly {\it et al.\/}]%
{SIMON J. LILLY$^1$, STEPHEN A. EALES$^2$, WALTER K. GEAR$^3$, 
TRACY M. WEBB$^1$, J. RICHARD BOND$^4$, LORETTA DUNNE$^2$}

\affiliation{$^1$Department of Astronomy, University of Toronto, 
60 St. George
Street, Toronto, Ontario M5S 3H8, Canada\\
[\affilskip]
$^2$Department of Physics and Astronomy, Cardiff University, P.O. Box 913,
Cardiff CF2 3YB, United Kingdom\\[\affilskip]
$^3$Mullard Space Science Laboratory, University College London, 
Holmbury St. Mary, Dorking, Surrey RH5 6NT, United Kingdom\\[\affilskip]
$^4$Canadian Institute for Theoretical Astrophysics, University of Toronto,
60 St. George Street, Toronto, Ontario M5S 3H8, Canada}

\begin{document}
\ifnfssone
\else
  \ifnfsstwo
  \else
    \ifoldfss
      \let\mathcal\cal
      \let\mathrm\rm
      \let\mathsf\sf
    \fi
  \fi
\fi

\maketitle

\begin{abstract}
Deep surveys of the sky at millimeter wavelengths have revealed a
population of ultra-luminous infrared galaxies (ULIRGs) at high
redshifts.  These appear similar to local objects of similar luminosities
(such as Arp220) but are much more ``important'' at high redshift than
at low reshift, in the sense that they represent a much larger fraction of the total
luminous output of the distant Universe than they do locally.  In fact the ULIRGs at high
redshift are producing a significant fraction ($\ge 15\%$) of the total
luminous output of the Universe averaged over all wavelengths and
all epochs.  The high $z$ ULIRGs could plausibly be responsible for
producing the metal-rich spheroidal components of galaxies,
including the bulges that are the subject of this conference. In
this case we would infer from the redshift distribution of the 
sources that much of this activity is probably happening
relatively recently at $z \le 2$.
\end{abstract}

\firstsection 
\section{Introduction}

     Despite a great deal of progress in recent years, there still
remain major uncertainties in our observational picture of the
formation and evolution of galaxies in the high redshift Universe.
Not least, the relationship between the star-formation activity
seen at high redshift and the present-day morphological components
of the galaxy population, including the bulges that are the
subject of this conference, remains unclear. The origin of the
stars in the metal-rich spheroidal components of present-day
galaxies, which constitute a half to two-thirds of all stars in
the Universe (see Fukugita et al 1998), is thus an unsolved
observational question. The formation of the bulk of metal-rich
spheroid stars in highly dissipational mergers of gas-rich systems
at high redshifts is an attractive scenario, except for the 
absence (hitherto) of a
substantial population of luminous star-forming galaxies at high
redshifts with the high star formation rates (several $10^2-10^3$
M$_{\odot}$yr$^{-1}$) that would be required to produce substantial spheroidal
components of galaxies on typical dynamical timescales of $10^8$ yr.
 
Several papers at this conference have highlighted the
evidence in the present-day Universe that the spheroidal
populations probably formed within the first 1/3 of the history of
the Universe, i.e. at $z \ge 1$.  Certainly, the evolution seen in the
optically-selected galaxy population out to $z \sim 1$ appears to be
primarily due to relatively small galaxies with irregular
morphologies and to the disk components of larger galaxies (see
e.g. Brinchmann et al 1998 and Lilly et al 1998a, also Guzman et
al 1997, Mallen-Ornelas et al, in preparation, and references therein) and it
is thus likely that the spheroids were to a large degree in place
by $z \sim 1$. The nature of the ultraviolet-selected ``Lyman-break''
galaxies seen at
$z > 3$ (Steidel et al 1996) and their relationship to present-day
galaxies is still quite uncertain (see e.g. Dickinson 1999, Trager
et al 1997 and references therein), and very little is really
known about the nature of galaxies in the crucial intermediate
redshift range $1.5 < z < 3$.

     However, it is very clear that the observational picture of
the high redshift Universe that has been gained from optical and
near-infrared observations must be seriously incomplete. The $\nu I_{\nu}$
energy content of far-IR/sub-mm background detected by the FIRAS and
DIRBE instruments on COBE (Puget et al 1996, Hauser et al 1998,
Fixsen et al 1998) is at least as large (see
e.g. Dwek et al 1998) as that of the optical/near-IR background
that is obtained by integrating the galaxy number counts (e.g. Pozzetti et
al 1998).  While some
of the far-IR background may result from AGN activity, it is
likely that of order a half of the energy from stellar
nucleosynthesis at cosmological redshifts emerges as re-processed
radiation in the far-IR. Indeed, in terms of the energy from
recent star-formation activity, the balance may be tipped even
further in favour of the far-IR because we know that a significant
fraction of the optical background will be coming from old stars –
- the energy of the optical/near-IR
background is already three times higher at K than at U,
see Pozzetti et al (1998).
 
    Determining the nature and redshifts of the sources
responsible for the far-IR/sub-mm background is therefore
vital to our understanding of galaxy evolution. Several groups
(e.g. Smail et al 1997, 1998, Hughes et al 1998, Barger et al 1998
and ourselves) are pursuing deep surveys in the sub-millimetre
waveband at 850 $\mu$m with the new SCUBA bolometer array
(Holland et al 1998, Gear et al in preparation) on the 15m James
Clerk Maxwell Telescope (JCMT) located on Mauna Kea.  Working at
850 $\mu$m has a number of rather interesting features since it is
well beyond the peak of the far-IR background ($100-200 \mu$m). Not
least, the $k$-corrections at 850 $\mu$m are extremely beneficial
as the rest-wavelength moves up with redshift towards the peak
of thermal dust emission around 100 $\mu$m. In consequence, a typical
star-burst galaxy (i.e. with an effective dust temperature of
around 30K and effective emissivity $\propto \nu^{1.5}$) has a
roughly constant observed flux density at 850 $\mu$m over the entire
$0.5 < z < 5$ redshift range, especially if $\Omega = 1$ (see Fig 3 of
Lilly et al 1999) and observations at 850 $\mu$m are thus as sensitive to
obscured star-formation at very high redshifts, $z \sim 5$ as they are
at $z \sim 0.5$!  This remarkable fact has a number of interesting
consequences. First, ``flux-density limited'' samples will
approximate ``luminosity limited'' (or ``volume limited'') samples;
secondly, the redshift distribution is likely to be only a weak
function of flux density; thirdly, the knowledge of precise
redshifts is not critical for determining bolometric luminosities;
and finally, one finds that the intuition of optical observers
towards quantities such as the redshift distribution sometimes
requires modification!

     In this paper, we review what is currently known about the
sources responsible for the 850 $\mu$m background.  We take $H_0 = 50
h_{50}$ kms$^{-1}$Mpc$^{-1}$
and for simplicity generally assume a matter-dominated
$\Omega = 1$ cosmology.

\section{Resolving the sub-mm background into discrete sources}

     In the last six months, four independent groups have
published first results from deep surveys at 850 $\mu$m. Smail et al
(1997,1998) have undertaken an ingenious survey using the
gravitational lensing effect of moderate redshift clusters of
galaxies to amplify background sub-mm sources and now have a
sample of 17 sources at $S_{850} > 6$ mJy (3$\sigma$). The remaining surveys
have been ``field'' surveys. Hughes et al (1998) published a single
very deep image of the HDF that revealed 5 sources at $S_{850} > 2$ mJy
(4$\sigma$), Barger et al (1998) had 2 sources at $S_{850} > 3$ mJy (3$\sigma$) and
our own program (Eales et al 1999) has 12 published sources
published with $S_{850} > 3$ mJy (3$\sigma$) with another 20 or so sources at
various stages of identification and analysis - the properties of these
appear consistent
with the first 12, but will not be discussed here.

     Given the small numbers involved, the number counts of
sources from these surveys are consistent (see Fig 1 - adapted
from Blain et al 1998b) and indicate substantial excesses over the
number of sources predicted in ``no-evolution'' replications of the
local IRAS 60 mm luminosity function (see Smail et al 1997, Eales
et al 1999).  The direct counts at $S_{850} > 2$ mJy have been extended
to about 1 mJy with a $P(D)$ analysis in the HDF (Hughes et al 1998)
and by a lens inversion analysis by Smail et al (1999).

\begin{figure} 
  \vspace{26.5pc}
  \caption{Lower panel: the cumulative number counts at 850 $\mu$m from recent published surveys
compared with a ``no-evolution'' model prediction (dotted line) from Eales et al (1999)
(adapted from Blain et al 1998b). Sources of data are: Solid dots (Eales et al 1999),
open square (Smail et al 1997), open circle (Barger et al 1998), 
stars (Hughest et al 1998), cross (Blain et al 1998). Upper panel:
the derived cumulative fraction of the 850 $\mu$m background (Fixsen et al 1998) that is produced 
by these sources assuming the solid curve in the lower panel.
} 
\end{figure} 

     While many of the sources have been detected at low S/N
ratios, the chopping and nodding employed in sub-mm observations
lend themselves to a number of straightforward statistical tests
(e.g. searching for negative images at the same level of
significance) and the great majority of the claimed sources are
probably real.  It should be noted that all of the blank-field
surveys are approaching or have reached the confusion limit. For
instance, at the $S_{850} \sim 3$ mJy 3$\sigma$ limit of our own survey there
are already about 40 beams per source, the conventional point at
which confusion effects become significant.

     The upper panel in Fig 1 shows the cumulative fraction of the
850 $\mu$m background (taken to be 12 mJy arcmin$^{-2}$ , Fixsen et al 1998) that is
produced by these detected sources. It should be noted that
already by 3 mJy we are accounting for 20\% of the background, a
fraction that increases to 50\% at the faintest limit, $S_{850} \sim 1$ mJy, probed by
the statistical studies.

\section{Identifications of the sub-mm sources}

\subsection{How reliable are the identifications}

     The SCUBA beam at 850 $\mu$m is 15 arcsec FWHM, necessitating a
probabilistic approach to identifications on deep optical or near-
infrared images. Some of the sub-mm sources are $\mu$Jy radio sources
enabling more accurate, arcsec-level, positions to be determined
and these can be identified relatively unambiguously. The fraction
of sources that are detectable as faint radio sources is not well
determined at this point. In the HDF, Richards (1999) claimed 3 of
5 of the Hughes et al sub-mm sources were detected at $S_{8.5GHz} > 10$
mJy (although this required a quite large and controversial offset of 6 arcsec
with respect to the Hughes et al (1998) sub-mm astrometric
reference frame). In our own CFRS-14 sample, for which the radio
catalogue extends to $S_{5GHz} \sim 16$ mJy (Fomalont et al 1992) we find
about 33\% of sub-mm sources to be radio sources (and also, since
they have similar surface densities, a similar fraction of radio
sources to be sub-mm sources).  In the future, millimetre
wavelength interferometry may produce better positions for the
remainder.

     All of the survey programs have searched for identifications
with extragalactic objects. It is possible to compute the
probability that the nearest member of a population of candidate identifications
(i.e. optical galaxies) with surface density $n$ is located within a
distance $d$ from a random position on the sky, $P = e^{- \pi nd^2}$
(e.g. Downes et al 1986) and this $P$ statistic has been used by
many workers in the identification of sub-mm sources (e.g. Hughes et al 1998,
Smail et al 1998).  There is already a subtlety in the use of $P$, in that if
the density of sources $n$ used is based on the magnitude of the
candidate identifications, i.e. $n(<m)$, then $P$ will suffer an {\it a posteriori}
bias, but this can be (and has been) dealt with either
analytically or through Monte Carlo simulations. The $P$ statistic
represents a starting point, but is not what is really required,
which is rather the probability that a particular claimed
identification is, in fact, correct. The quantity $P$ tells us the
fraction of sources in a sample of size $N$ that would be expected
to have an incorrect candidate identification lying within this
distance $d$, i.e. $N_{spurious}(<P) = NP$. Thus, a low value of $P$ for
any individual source is not, on its own, enough to make
an identification secure. Rather, one has to look at the sample as
a whole and determine the number of identifications in the sample
(with a certain value of $P$) relative to the number of spurious
identifications (with that same $P$) that would have been expected
if the two populations were completely unrelated.
Only if this ratio is high can a particular individual source with
that value of $P$ be regarded as securely identified. This is
illustrated in Fig 2, which shows the distribution of (corrected)
$P$ values for the identifications in the three main published
programs (the solid histograms) compared with the distribution of
$P$ expected if the optical and sub-mm populations were completely
unrelated (the smooth line).  Statistically, 
sources lying ``below'' the smooth
line cannot therefore be regarded as identified, regardless of
their value of $P$, since that number of objects would have been
expected by chance!

\begin{figure} 
  \vspace{26.5pc}
  \caption{The distribution of $P$ values in the three main published surveys (solid histograms -
upper two panels Hughes et al (1998) based on magnitudes and redshifts, then
our own sample and the Smail et al (1998) lensing sample). The best
measure of whether individual identifications are correct is given by comparing the number of identifications in the whole sample with a particular value of $P$ with the number,$NP$, that would have been expected 
by chance (solid line). This suggests that about 50\% of the sources in all the samples
are have been correctly identified.} 
\end{figure} 

     Our conclusion from Fig. 2 is that in all of the deep sub-
mm samples studied to date we have reliable identifications for
only about half (40-60\%) of the sub-mm sources, regardless of
whether identifications for the remainder have been claimed or
not.  This is a handicap, but as we will see below, it is not as
serious as one might suppose.  Furthermore, the statistics of the
identifications already enable us to make an important statement:
{\it At least half (and quite possibly all) of the high latitude 850 $\mu$m
sources must be extragalactic in nature}.  

\subsection{The redshift distribution of the identifications}

     Hughes et al in their HDF sample emphasized the high
redshifts of their identifications. In our own  program (Lilly et
al 1998, 1999), we found that many of the sub-mm identifications
had already been catalogued in the CFRS program and that in fact
three of the first 12 sources had spectroscopic redshift
measurements at $z < 1$ (at $z$ = 0.074, 0.55 and 0.66). For the
remainder, we have estimated redshifts on the basis of the optical-
infrared $UVIK$ colours. We concluded that all the initial eight identifications
(of which at least six may be regarded as secure) were optically
luminous galaxies (comparable to present-day $L*$) spanning a broad
range of redshifts $0.08 < z < 3$, with four at $z < 1$.  The
upper limit at $z \sim 3$ comes from detections in the $U$-band.  
With the present rather limited
data, the observed properties of the four unidentified empty
field sources in our sample would be broadly consistent with those
of the identified galaxies if they were placed anywhere over a
rather wide range of redshifts, $2 < z < 10$. Redshifts as low as $z
\sim 1$ however would not be excluded by the present data but would
require a higher extinction, as in VIIZw031 (see Trentham et al 1999)
A reasonable guess for the median redshift is $<z> \sim 2$.  

     As discussed in Lilly et al (1999 - see their Fig 10ab), these results appear to be broadly
similar to those of the other surveys, especially if the
Richards (1999) modification of the HDF identifications are adopted. The lensed sample of
Smail et al (1998) does not at present have redshift estimates (except for
constraints based on detection in $B$ or $V$) but appears to have a
similar distribution in $I_{AB}$ magnitude especially when an average
lens amplification of a factor of 2.5 is taken into account.

\subsection{The nature of the sub-mm sources}

	As noted above, any source detected at $S_{850} \ge 3$ mJy that lies
at $z > 0.5$ must have a luminosity above that of Arp 220, i.e.
$L \ge 3 \times 10^{12} h_{50}^{-2} L_{\odot}$. Assuming the energy comes from
star-formation as opposed to black-hole accretion, this 
luminosity corresponds to a 
substantial star-formation rate of $\ge 600 h_{50}^{-2} M_{\odot}$ yr$^{-1}$.
     
The broad-band spectral energy distributions of the
identifications in our own sample, as defined from the optical
through the far-IR component to the radio, from measurements or
limits at 0.8 $\mu$m, 15 $\mu$m, 450 $\mu$m, 850 $\mu$m and at 5 GHz, are
consistent with the measured/estimated redshifts of the
identifications and a rest-frame SED that broadly matches that of
Arp 220.

     The galaxies have a range of optical colours, but are on
average a little redder in ($V-I$) than typical field galaxies,
consistent with what is known about the ultraviolet properties of
local ULIRGs (Trentham et al 1999).  The ($V-I$),$I$
 
for our identifications and for those in the Smail et al program match nicely
the expectations based on local ULIRGs (Trentham et al 1999).
The HST morphologies of the $z > 0.5$ identifications in our sample
range from relatively normal looking galaxies to clear examples of
mergers, but nearly all show some sign of peculiarity in the form
of secondary nuclei or asymmetrical outer isophotes.  Little is
known about the ultraviolet morphologies of ULIRGs at low
redshift, but the Trentham et al (1999) study shows considerable 
diversity and substantial
differences from the optical morphologies.

In summary,{\it in essentially all respects that can presently
be studied, the $z > 0.5$ sources in our sample appear to be very
similar to local ULIRG prototypes such as Arp220}.

\section{The significance of ULIRGs at high redshift}
 
     The results outlined above lead robustly to a very important
conclusion: {\it ULIRGs as a class are a much more important component
of the galaxy population (in that they produce a much higher
fraction of the total luminous output) at high redshift than at
low redshift}.  In the local Universe, ULIRGs of luminosities
greater or equal to that of Arp 220 (i.e. $2 \times 10^{12} h_{50}^{-2} L_{\odot}$)
contribute only about 1\% of the far-IR luminous output of the galaxy
population (Soifer et al 1987, Saunders et al 1990, 
Sanders and Mirabel 1996).  In contrast, at
high redshifts ($z > 1$), similar objects must produce at least 30\%
of the far-IR/sub-mm background (for $\Omega = 1$, more for 
low $\Omega$ since Arp220 would lie further down the $N(S)$ 
distribution), which we have seen
is at least equal in energy content to the optical/near-IR
background.  Thus, high luminosity obscured objects are much more
common, relatively, at high redshift, and are in fact producing a
substantial fraction (15\%) of the total luminous output of the Universe
averaged over all epochs and all wavebands.

     The cumulative bolometric luminosity function in the far-IR 
derived from our own sample
is shown in Fig 3 compared with the local IRAS luminosity
function and the ultraviolet luminosity function for the Lyman-break 
galaxies constructed by Dickinson (1999).

\begin{figure} 
  \vspace{26.5pc}
  \caption{Estimate of the cumulative bolometric luminosity function
of the sub-mm population from our own sample (points), the local IRAS 60$\mu$m population,
and the $z \sim 3$ ultraviolet-selected ``Lyman-break'' population (uncorrected
for extinction). The $1 < z < 3$ sub-mm points show minimum and maximum values
according to whether the four ``empty field'' sources are at $z > 3$ or $1 < z < 3$.
Likewise, the $3 < z < 8$ point assumes that the ``empty fields'' are at $z > 3$.} 
\end{figure} 

\section{Interpretation of the redshift distributions}

     The relatively small number of sources (no more than 50\% of
the sample) that can possibly be at very high redshifts ($z > 3$)
already sets quite strong constraints on the amount of high
luminosity obscured star-formation that can take place at these
redshifts.  This is because, as pointed out in previous papers
(Lilly et al 1998, 1999), the beneficial $k$-corrections
produce a strong weighting of high
redshift star-formation activity in the production (in redshift
space) of the 850 $\mu$m background relative to the production of
stars (in redshift space). This weighting is simply
$f_{\nu}(\nu_{em})/f_{\nu}(\nu_{obs})$, or $(1+z)^{3.5}$ over much of the redshift range of
interest $0 < z < 6$.  In Fig 4 (from Lilly et al 1999), 
we have computed the redshift distribution of the
850 $\mu$m background light for a number of different star-formation
histories, assuming that the energy of this star-formation emerges
with the spectral energy distribution of an obscured star-burst,
like Arp 220. It can be seen that galaxy formation/evolution scenarios in
which 50\% of all dust enshrouded star-formation in the Universe
occurred prior to $z = 3$ predict that 85\% of the 850 $\mu$m
background had been produced at $z > 3$.  
The distribution of observed light at
$S_{850} > 2.8$ mJy in our identified source sample is also shown. This
does not reach unity because the unidentified sources have been
omitted since their redshifts are unconstrained (but it is assumed for
this purpose that they have $z > 2.5$) and the contribution from
the two less securely identified galaxies estimated to lie at
around $z ~ 2.5$ is shown as a dotted line.

\begin{figure} 
  \vspace{26.5pc}
  \caption{ Production of the 850 $\mu$m background from different star formation histories: 
The left hand panel shows five different heuristic star-formation histories. For each model, the 
cumulative production of stars is shown in the center panel, and the cumulative distribution 
of light in the 850 $\mu$m background is shown in the right-hand panel. Because of the highly 
beneficial $k$-corrections at 850 $\mu$m, the light in the background is heavily weighted in favour of 
high redshift star-formation.  Models in which half the obscured star-formation in the Universe 
occurred prior to $z \sim 3$ 
predict that only 15
all of the sources with $S_{850} < 3$ mJy have z > 3 - a highly unlikely situation.  
The irregular line in the left-most panel shows the distribution in redshift of the background 
produced by the observed sources - illustrating the effect of assuming that the sources below 
the limit of the survey in fact have the same redshift distribution. 
This would require a falling
luminosity density at high redshifts would imply that most stars formed in these obscured objects did so
at $z \le 2$. 
} 
\end{figure} 

Even if we
make no assumption at all about the redshifts of the fainter
sources with $S_{850} < 3$ mJy, our observations would already appear
to require that at least 15\% of the 850 $\mu$m background must be
produced at $z < 3$, which is only barely consistent with a scenario
in which 50\% of obscured star-formation takes place at $z \ge 3$.

If we speculatively assume that the
redshift distribution of fainter sources with $S_{850} < 3$ mJy follows
that at $S_{850} > 3$ mJy (a plausible, but not watertight, assumption
given the flatness of the 850 $\mu$m flux density-redshift relation,
see also the models of Blain et al 1998a)
then our results then suggest that the great bulk of obscured star-
formation in the Universe occurred at redshifts $z < 3.0$.  While
this analysis can not be regarded as conclusive until we penetrate
deeper in to the background, Fig 4 suggests that the cumulative
production of the 850 $\mu$m background appears to follow well the
expectations of models in which the luminosity density in the 
far-IR (at least in high luminosity obscured objects) peaks in the $1.2
< z < 2$ range and falls thereafter.  Interestingly, initial
indications for a decline in the ultraviolet luminosity density of the Universe at
high redshifts (Madau 1996, 1997) have not been borne out by more
recent work (Steidel et al 1999).

\section{The relationship to the Lyman-break ultraviolet-selected galaxy
population}

     As shown in Fig 12 of Lilly et al (1999) the bolometric output
in the far-IR of the high luminosity ($L \ge 3 \times 10^{12} h_{50}^{-2} L_{\odot}$)
high $z$ ULIRG population already matches that
in the ultraviolet of the whole ``Lyman-break'' population of
galaxies, even though the former only comprise the ``top''
20\% of the 850 $\mu$m background.

     Estimates of the reddening of the Lyman-break galaxies based
on the observed ultraviolet continuum slope suggest that for
typical LBG the far-IR luminosity would be between $2-7$ times that
seen in the ultraviolet (Dickinson 1999, see also Pettini et al
1998) for SMC and Calzetti extinction curves, with higher
values being claimed by Meurer et al (1997).  These typical
objects, with ``corrected'' star-formation rates of up to 
$30-100 h_{50}^{-2} 
M_{\odot}$yr$^{-1}$
would be undetectable with SCUBA at present, but would have to be
responsible for a significant fraction of the background.

     Obviously the estimation of bolometric luminosities on the
basis of extinction in the ultraviolet is highly uncertain –
requiring an assumed extinction curve that largely reflects the
geometrical distribution of stars and dust. This is especially
true in the high extinction regime (see the three examples of
local ULIRGs in Trentham et al 1999).  Nevertheless, preliminary
indications (Steidel, private communication) are that the number
of very highly extinguished LBG with ``corrected'' Arp220-level
luminosities (i.e. after correction with the Calzetti reddening
curve) are roughly consistent with the number directly observed in
the sub-mm surveys described here, which have $\phi \sim 10^{-4}h_{50}^3$ Mpc$^{-3}$,
Fig 3).  This agreement is encouraging, though possibly fortuitous given the 
uncertainties in the reddening correction applied to the optical sample. 
The sub-mm sample may also contain
some ULIRGs that are so heavily obscured as to be completely
absent from the present LBG samples.

\section{The nature of the obscured energy sources}

     A difficult question concerns the fraction of the far-IR
energy that comes from hidden active galactic nuclei. In the local
Universe, the evidence from mid-IR emission features
(Genzel et al 1998) is that AGN provide a significant but not
dominant (25\% - 50\%)
contribution to ULIRGs at these luminosities and this
seems a reasonable first guess as to the situation at high
redshifts. The ultraviolet spectrum of the highly luminous sub-mm source
SMM02399-0136 (Ivison et al 1998) shows spectroscopic indications for an
AGN but this same source also
exhibits strong, star-burst like, CO detections 
(Frayer et al 1998, see also Frayer et al 1999). For all but the brightest sources (which may
be tackled by SIRTF) these mid-IR diagnostics may be unobservable
until NGST flies - and even then only if it has a mid-IR
spectroscopic capability. 

Ascribing a dominant fraction of the
energy output of this population to AGN would require a major
upwards revision in the total energy output of AGN.  On the other hand,
several authors have stressed the inadequacy of the ``known'' quasar
population to produce the required mass of black holes (integrated
over the population). Using the Magorrian et al (1996) relationship between black hole mass and stellar 
bulge mass ($M_{BH} \sim 0.006 M*$) in local galaxies and
assuming a radiative efficiency $\epsilon \sim 0.1$ for black hole accretion
and an energy release of 
$0.016M_Zc^2$ for the return of $m_Z$ of metals (Songaila et al 1990)) it is easy 
to show that the bolometric light output associated the production of the black holes and 
stellar metals in typical spheroids should be comparable:

\bigskip

$\frac{L_{AGN}}{L_{star}} \sim \frac{\epsilon M_{BH} c^2}{0.016 M_Z c^2} \sim \frac{M_{BH}}{M_{star}} \frac{1}{0.0032} \sim 2$
  
\bigskip

On the other hand, the approximate consistency with the 
extinction-corrected properties of the LBG population noted 
above suggests that much of the far-IR background is indeed 
coming from stars.  Observational resolution of this important 
question at better than the factor of two level will be challenging.

\section{The formation of spheroids?}

The identification, in the sub-mm, of a population of galaxies at 
high redshift that are producing a substantial fraction of present 
day stars in high luminosity systems is important because it is then 
attractive to identify these as producing the metal-rich spheroidal components 
of galaxies, including the bulges of present-day spiral galaxies. 
Local ultra-luminous IR galaxies have long been proposed 
as being triggered by major mergers and resulting in the production of 
massive spheroids (see Sanders and Mirabel 1997 and references therein). 
The high individual luminosities ($\ge 3 \times 10^{12} h_{50}^{-2} L_{\odot}$) 
and implied star-formation rates 
($\ge 600 h_{50}^{-2} M_{\odot}$yr$^{-1}$) are 
consistent with making substantial stellar populations on dynamical 
timescales. 

It should be stressed that the ULIRG population revealed in the sub-mm 
surveys at high redshift has a sufficiently high number density (more 
than 100 times higher than in the present-day Universe - Fig 3) that 
they could be responsible for producing a significant fraction of all 
stars that have been formed in the Universe, since they are responsible 
for producing, in the far-IR, a significant fraction of the entire 
bolometric extragalactic background light.  In more
absolute terms, a
star-formation rate of $600 h_{50}^{-2} M_{\odot}$yr$^{-1}$ maintained 
for $4 h_{50}^{-1}$ Gyr 
at a number density of 
$\phi \sim 10^{-4}h_{50}^3$ Mpc$^{-3}$ would yield a stellar mass density of 
$2.4 \times 10^8 M_{\odot}$ Mpc$^{-3}$. It should be recalled (e.g. Fukugita et al 1998) 
that the spheroids contain a half to two-thirds of all stars in the Universe
or about $0.7 - 2.0 \times 10^8 h M_{\odot}$ Mpc$^{-3}$. So the numbers are in the
right ball-park.

The 
combination of the high integrated production of stars, the high 
star-formation rates, the incidence of merger-like morphologies and 
the obvious presence of substantial amounts of dust, make it attractive, 
though still speculative, to associate these galaxies with the production 
of the metal-rich spheroid component of galaxies.  In this case, these 
first data from our survey (see Fig. 4) suggest that much of this activity, 
conservatively at least 50\%, and probably much more, has happened 
at relatively recent epochs, i.e. $z < 3$.  

\begin{acknowledgments}
The research of SJL and JRB is supported by the Natural Sciences Engineering Research Council of Canada and by the Canadian Institute for Advanced Research. The Research of SAE and WKG is supported by the Particle Physics and Astronomy Research Council in the
United Kingdom. The support of all of these agencies is gratefully 
acknowledged.
\end{acknowledgments}


\begin{thebibliography}{} 

\bibitem[]{}{Barger, A., Cowie, L., Sanders, D., Fulton, E., Taniguchi, Y., Sato, Y., Kaware, K., Okuda, H., 1998, Nature, 394, 248.}

\bibitem[]{}{Blain, A., Longair, M., 1993, MNRAS, 264, 509.}

\bibitem[]{}{Blain, A., Smail, I., Ivison, R., Kneib, J.P., 1998, astro-ph/9806062.}

\bibitem[]{}{Blain, A., Kneib, J.P., Ivison, R., Smail, I., 1998, astro-ph/9812412.}

\bibitem[]{}{Brinchmann, J., Abraham, R., Schade, D., Tresse, L., Ellis, R., Lilly, S., Le Fèvre, O., Glazebrook, K., Hammer, F., Colless, M., Crampton, D., Broadhurst, T., 1998, ApJ, 499, 112. }

\bibitem[]{}{Connolly, A., Szalay, A., Dickinson, M., SubbaRao, M., Brunner, R., 1997, ApJ, 486, L11.}

\bibitem[]{}{Dickinson, M., 1999, In ``The Hubble Deep Field'' (ed. M. Livio, M. Fall and P.
Madau), p219}

\bibitem[]{}{Downes, A.J.B., Peacock, J.A., Savage, A., carrie, D., 1986, MRAS, 218, 31.}

\bibitem[]{}{Dwek, E., Arendt, R., Hauser, M., Fixsen, D., Kelsall, T., Leisawitz, D., Pei, Y., Wright, E., Mather, J., Moseley, S., Odegard, N., Shafer, R., Silveberg, R., Welland, I., 1998, ApJ, 508, 106.}

\bibitem[]{}{Eales, S., Lilly, S., Gear, W., Bond, J.R., Dunne, L., Hammer, F., Le Fèvre, O., Crampton D., 1998, ApJ, in press (Paper 1)}

\bibitem[]{}{Fixsen, D., Dwek, E., Mather, J., Bennet, C., Shafer, R., 1998, ApJ, 508, 123.}

\bibitem[]{}{Fomalont, E., Windhorst, R., Kristian, J., Kellerman, K., 1991, AJ, 102, 1258.}

\bibitem[]{}{Frayer, D.T., Ivison, R.J., Scoville, N.Z., Yun, M., Evans, A.S., Smail, I., Blain,A.W., Kneib, J.-P., ApJ 506, 7}

\bibitem[]{}{Frayer, D.T., Ivison, R.J., Scoville, N.Z., Evans, A.S., Yun, M., Smail, I.,
Barger, A., I., Blain,A.W., Kneib, J.-P., ApJLett 514, L13}  

\bibitem[]{}{Genzel, R., Lutz, D., Sturm, E., Egami, E., Kunze, D., Moorwood, A.F.M., Rigopoulou, D., Spoon, H., Sternberg, A., Tacconni-Garman, L., Tacconi, L., Thatte, N., 1998, ApJ, 498, 579.}

\bibitem[]{}{Guzman, R., Gallego, J., Koo, D., Phllips, A., Lowenthal, J., Faber, S.M., Illingworth, G., Vogt, N., 1998, ApJ, 489, 559.}

\bibitem[]{}{Haehnelt, M.G., Natarajan, P., Rees, M.J., 1998, MNRAS, 300, 817.}

\bibitem[]{}{Hauser, M., Arendt, R., Kelsall, T., Dwek, E., Odegard, N., Welland, J., Freundenreich, H., Reach, W., Silverberg, R., Modeley, S., Pei, Y., Lubin, P., Mather, J., Shafer, R., Smoot, G., Weiss, R., Wilkinson, D., Wright, E., 1998, ApJ 508, 25.}

\bibitem[]{}{Holland, W.S., Robson, E.I.,Gear, W.K., Cuningham, C.R., Lightfoot, J.F., Jenness, T., Ivison, R., Stevens, J.A., Ade, P.A.R., Griffin, M.J., Duncan, W.D., Murphy, J.A., Naylor, D.A., 1998, MNRAS, in press, astro-ph/9809122.}

\bibitem[]{}{Hughes, D., Serjeant, S., Dunlop, J., Rowan-Robinson, M., Blain, A., Mann, R., Ivison, R., Peacock, J., Efstathiou, A., Gear, W., Oliver, S., Lawrence, A., Longair, M., Goldschmidt, P., Jenness, T., 1998, Nature 394, 241. }

\bibitem[]{}{Ivison, R., Smail, I., Le Borgne, J-F., Blain, A., Kneib, J-P., Kerr, T., Bezecourt, J., Davie, J., 1998, MNRAS, 298, 583.}

\bibitem[]{}{Lilly, S., Cowie, L., 1987, In ``Infrared Astronomy with Arrays'' (eds. Wynn-Williams, G., Becklin, E.), UH, Honolulu, p.473.}

\bibitem[]{}{Lilly, S.J., Le Fevre, O., Crampton, D., Hammer, F., Tresse, L., 1995a, ApJ, 455, 50. }

\bibitem[]{}{Lilly, S.J., Hammer, F., Le Fèvre, O., Crampton, D., 1995b, ApJ, 455, 75. }

\bibitem[]{}{Lilly, S.J., Le Fèvre, O., Hammer, F., Crampton, D., 1996, ApJ, 460, L1. }

\bibitem[]{}{Lilly, S.J., Schade, D., Ellis, R., Le Fèvre, O., Brinchmann, J., Tresse, L., Abraham, R., Hammer, F., Crampton, D., Colless, M., Glazebrook, K., Mallen-Ornelas, G., Broadhurst, T., 1998a, ApJ, 500, 75.}

\bibitem[]{}{Lilly, S.J., Eales, S.A., Gear, W., Dunne, L., Bond, J.R., Hammer, F., Le Fèvre, O., Crampton, D., 1998b, In ``NGST: Scientific and Technical Challenges'', Proceedings of the 34th Liège International Astrophysics Colloquium, (ed. B. Kaldeich-Schürmann), ESA.}

\bibitem[]{}{Madau, P., Ferguson, H., Dickinson, M., Giavalisco, M., Steidel, C., Fruchter, A., 1996, MNRAS, 283, 1388.}

\bibitem[]{}{Madau, P., Pozetti, L., Dickinson, M., 1998, ApJ, 498, 106.}

\bibitem[]{}{Meurer, G., Heckman, T., Lehnert, M., Leitherer, C., Lowenthal, J., 1997, AJ, 114, 51}.

\bibitem[]{}{Pettini, M., Kellogg, M., Steidel, C., Dickinson, M., Adelberger, K., Giavalisco, M.,
1998, ApJ, 508, 539.}

\bibitem[]{}{Pozzetti, L., Madau, P., Zamorani, G., Ferguson, H.C., Bruzual, G., 1998, MNRAS, 298, 1133.}

\bibitem[]{}{Puget, J-L., Abergel, A., Bernard, J-P., Boulanger, F., Burton, W.B., Desert, F.X., Hartmann, D., 1996, A\&A, 308L, 5P.}

\bibitem[]{}{Richards, E.A., 1998, astro-ph/9811098}

\bibitem[]{}{Sanders, D., Mirabel, I., 1996, ARA\&A, 34, 749.}

\bibitem[]{}{Saunders, W., Rowan-Robinson, M., Lawrence, A., Efstathiou, G., Kaiser, N.,
Ellis, R.S., Frenk, C., 1990, MNRAS, 242, 318.}

\bibitem[]{}{Sawicki, M., Lin, H., Yee, H., 1997, AJ, 113, 1.}

\bibitem[]{}{Smail, I., Ivison, R., Blain, A., 1997, ApJ, 490, L5.}

\bibitem[]{}{Smail, I., Ivison, R., Blain, A., Kneib, J-P, 1998, ApJ, 507, L21.}

\bibitem[]{}{Songaila, A., Cowie, L.L., Lilly, S.J., 1990, ApJ, 348, 371.}

\bibitem[]{}{Steidel, C., Giavalisco, M., Pettini, M., Dickinson, M., Adelberger, K.,
1996, ApJLett 462 L17.}

\bibitem[]{}{Steidel, CV., Adelberger, K., Giavalisco, M., Dickinson, M, Pettini, M.,
1998, asyro-ph/9811400.}

\bibitem[]{}{Trager, S., Faber, S., Dressler, A., Oemler, A., 1997, ApJ, 485, 92}

\bibitem[]{}{Trentham, N., Kormenday, J., Sanders, D., 1999, 1999, astro-ph/9901382.}


\end{thebibliography}
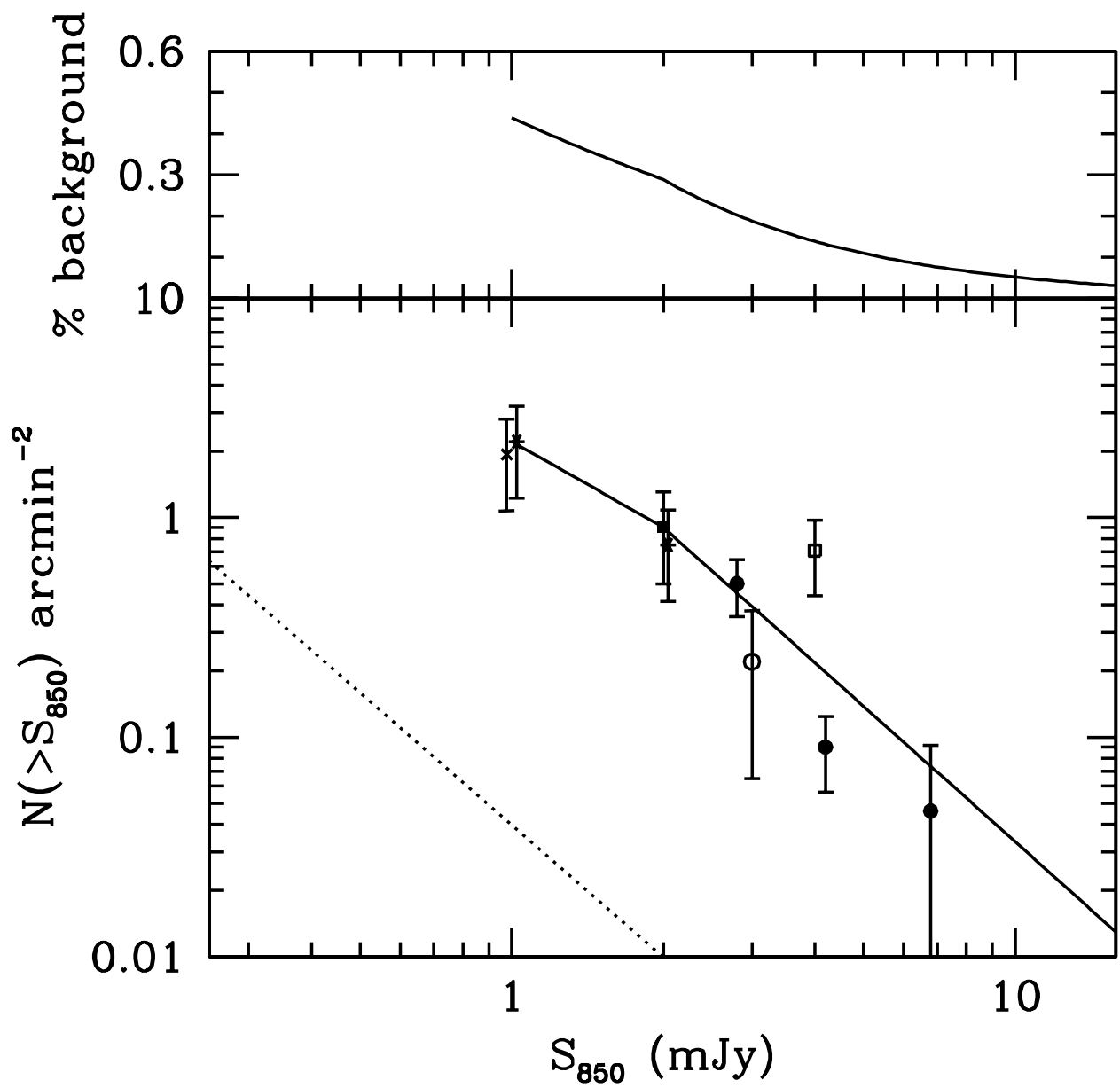
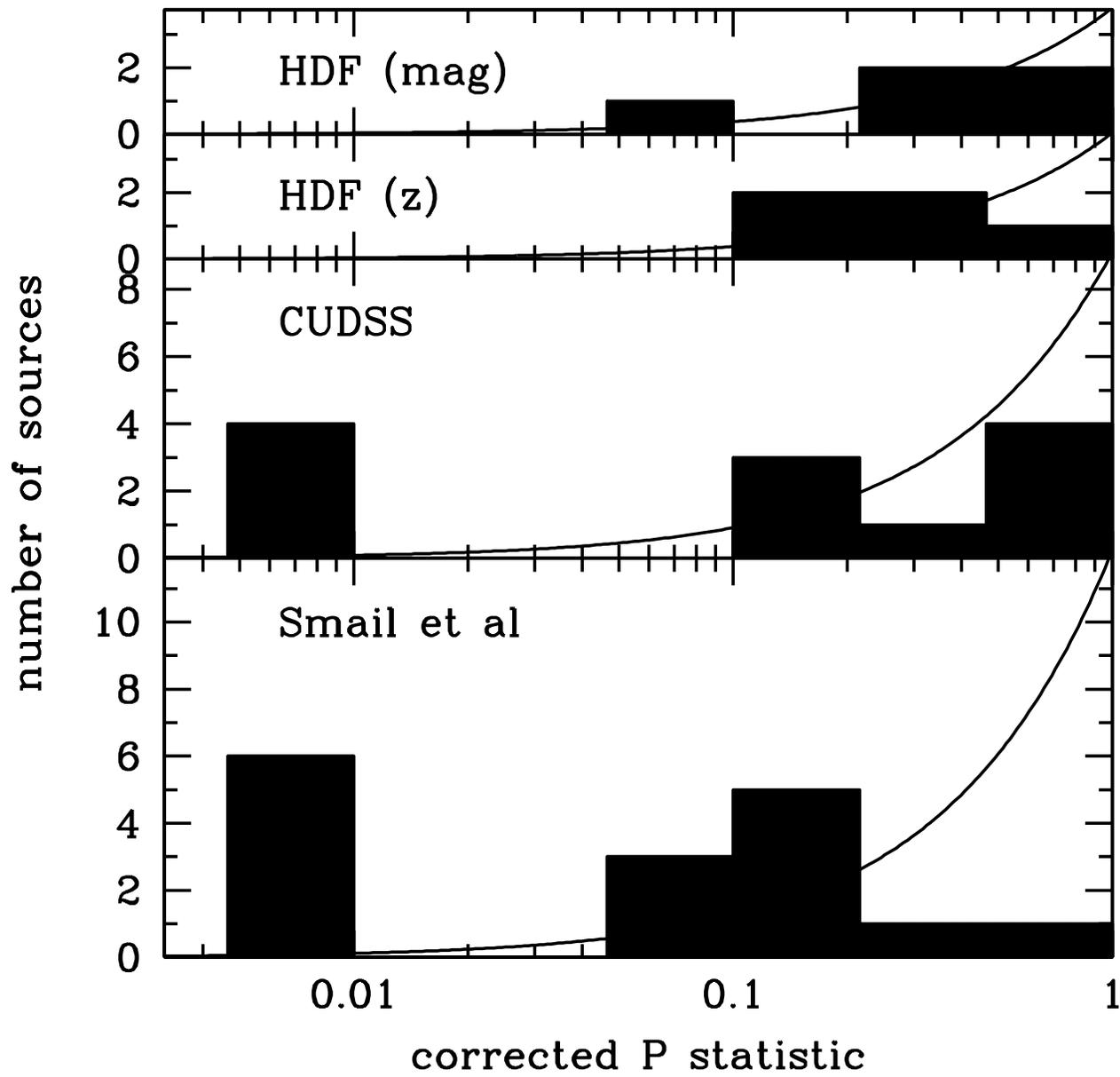
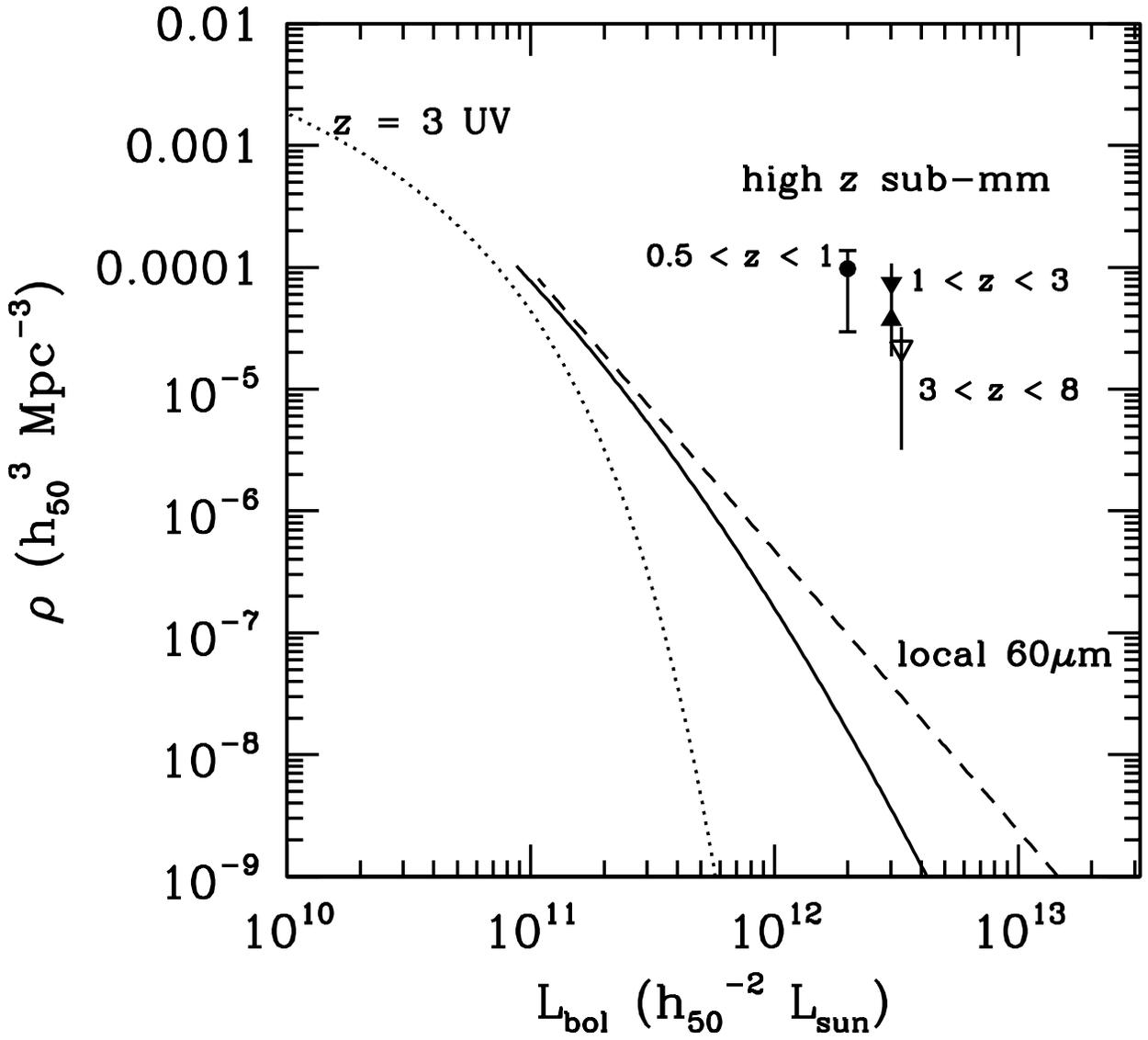
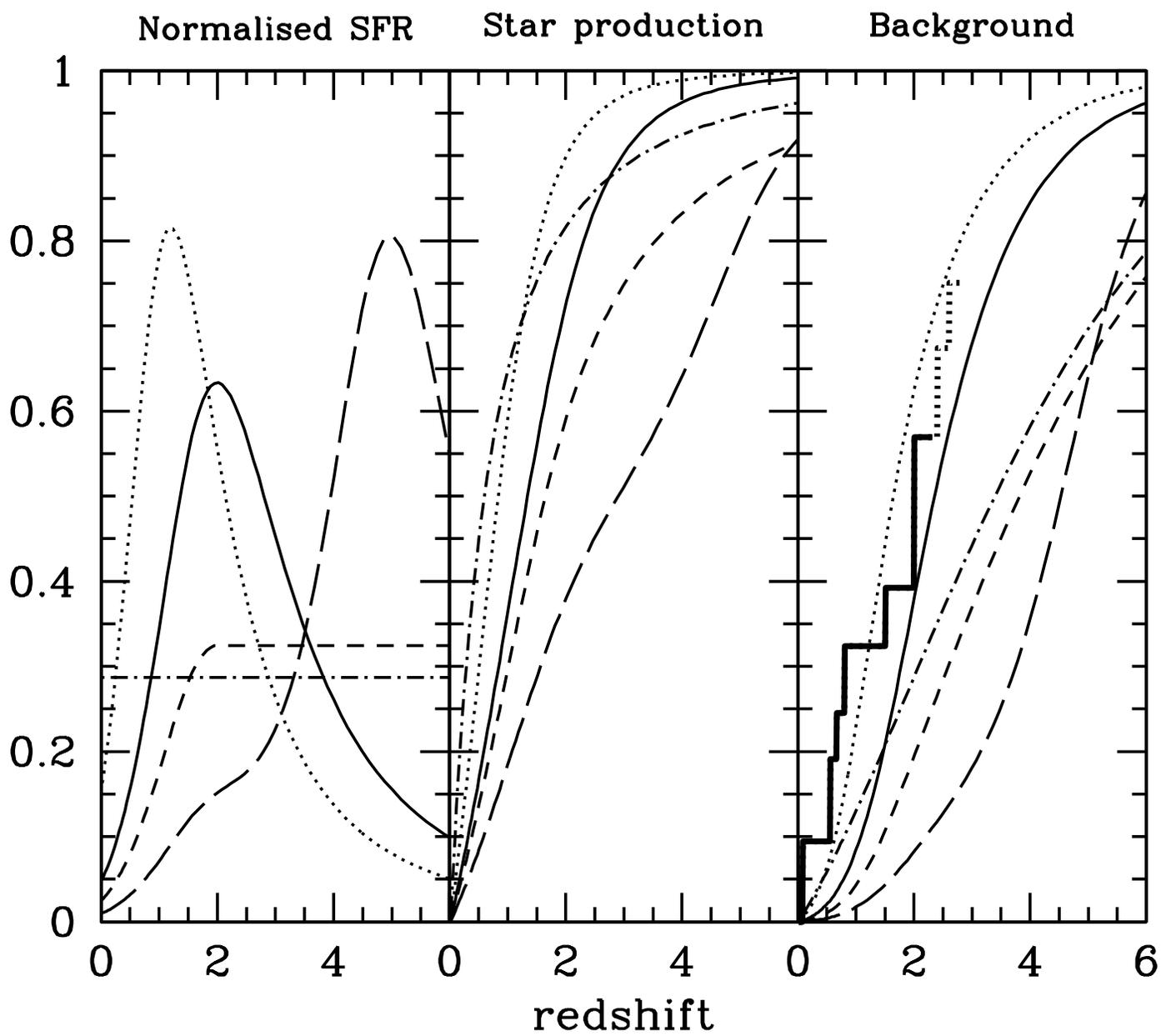
\end{document}